# Diffusion-controlled Alloying of Single-phase Multi-principal Transition Metal Carbides with High Toughness and Low Thermal Diffusivity

Chong Peng[1,2,a], Xiang Gao[2,a], Mingzhi Wang[1,a,b], Lailei Wu[1], Hu Tang[1,2], Xiaoming Li[3], Qian Zhang[2], Yang Ren[4], Fuxiang Zhang[5], Yuhui Wang[6], Bing Zhang[1], Bo Gao[2], Qin Zou[1], Yucheng Zhao[1], Qian Yang[1], Dongxia Tian[1], Hong Xiao[2], Huiyang Gou[2,b], Wenge Yang[2], Xuedong Bai[3], Wendy L. Mao[7] and Ho-kwang Mao[2,8]

[1]*State Key Laboratory of Metastable Materials Science and Technology, Yanshan University, Qinhuangdao 066004, People's Republic of China;*

[2]*Center for High Pressure Science and Technology Advanced Research, Beijing 100190, People's Republic of China;*

[3]*Beijing National Laboratory for Condensed Matter of Physics and Institute of Physics, Chinese Academy of Sciences, Beijing 100190, People's Republic of China;*

[4]*Advanced Photon Source, Argonne National Laboratory, Argonne, Illinois 60439, USA;*

[5]*Division of Materials Science & Technology, Oak Ridge National Laboratory, Oak Ridge TN 37831, USA;*

[6]*National Engineering Research Center for Equipment and Technology of Cold Strip Rolling, Yanshan University, Qinhuangdao 066004, China;*

[7]*Geological Sciences, Stanford University, Stanford, CA 94305;*

[8]*Geophysical Laboratory, Carnegie Institution of Washington, Washington, DC 20015, USA*

[a]C.P., X.G. and M.Z.W. contributed equally to this work.

[b]Authors to whom correspondence should be addressed. Electronic addresses: wmzw@ysu.edu.cn(M.Z.W.); huiyang.gou@hpstar.ac.cn(H.Y.G.)

**Abstract**

Multicomponent alloying has displayed extraordinary potential for producing exceptional structural and functional materials. However, the synthesis of single-phase, multi-principal covalent compounds remains a challenge. Here we present a diffusion-controlled alloying strategy for the successful realization of covalent multi-principal transition metal carbides (MPTMCs) with a single face-centered cubic (FCC) phase. The increased interfacial diffusion promoted by the addition of a nonstoichiometric compound leads to rapid formation of the new single phase at much lower sintering temperature. Direct atomic-level observations via scanning transmission electron microscopy demonstrate that MPTMCs are composed of a single phase with a random distribution of all cations, which holds the key to the unique combinations of improved fracture toughness, superior Vickers hardness, and extremely lower thermal diffusivity achieved in MPTMCs. The present discovery provides a promising approach toward the design and synthesis of next-generation high-performance materials.

## I. INTRODUCTION

The discovery and design of novel materials with desirable properties represents one of the ultimate objectives for materials science and technology. To achieve this goal, the general route for materials scientists is effective manipulations of composition and microstructure of materials, nevertheless, small changes in composition and microstructure could have entirely different physical properties and mechanical performance. Therefore, the exploration of highly specialized approaches to optimize them is urgent for the increasing needs of the high-performance engineering materials. The development of next generation materials that exhibit a combination of high strength, ductility, and fracture toughness is vital for realizing advanced high-performance structural engineering materials. Unfortunately, the intrinsic brittleness of many existing structural materials greatly limits their practical applications.[1-6] Attempts have been made to develop advanced engineering materials with both enhanced strength and ductility through the fabrication of novel multi-principal crystalline compounds or composite materials.[7-9] Extensive experimental studies have demonstrated that the specific combinations of two or more components in composite materials endow them with exceptional mechanical performance, which is inaccessible in monolithic materials.[1-9] To date, however, the pathways toward new, multi-principal crystalline phases achieving simultaneous strength and ductility have been limited.

High-entropy alloying represents an effective strategy for the selection and design of advanced engineering and functional materials exceeding conventional alloys. In high entropy alloys (HEAs), the formation of unique, single-phase microstructures with chemical and structural disorder leads to their excellent mechanical performance, including improved strength and damage tolerance, high thermal stability, and good wear resistance, as well as excellent electrical and thermal properties.[10-20] Significant efforts have been devoted to exploring complex composition space and microstructure in search of more attractive structural and functional properties of HEAs. Although the high-entropy alloying has been successfully applied to the combinations of individual metal elements,[10-20] it remains elusive for the single-phase formation of multiple compounds.

Refractory transition metal carbides and nitrides are of interest for a wide range of applications because of their attractive chemical and physical properties such as high thermal conductivity, extreme hardness, and low chemical reactivity and work functions.[21] As excellent high-temperature engineering materials, however, transition metal carbides are very difficult to sinter and machine with conventional methods due to their low self-diffusion coefficients and strong covalent bonding.[22-24] Assuming the metal cations in transition metal carbides can be replaced arbitrarily without breaking the original symmetry, TiC, TaC, ZrC, WC, TiN and others can be sintered into a single phase at low or moderate temperatures, and the excellent mechanical (both high hardness and toughness) performance can be maintained simultaneously. The recent discovery of five-component metal oxides[25] and diborides[26] has enlarged the possibilities for synthesis of multi-principal materials from combinations of compounds, offering untapped opportunities to conduct large-scale materials design. However, there are no reported routes for preparing a single-phase from several transition metal carbides and nitrides at relatively low temperature. Here we report a diffusion controllable synthetic procedure for the fabrication of multi-principal covalent transition metal carbides with equal amounts at much lower sintering temperature. We exploit the phase formation and sintering mechanism for three to seven components by microstructure analysis and direct atomic-level observations, and these phases are found to exhibit unique combinations of improved fracture toughness, superior Vickers hardness and extremely low thermal diffusivity. The alloying technique developed here has broad implications for the rational design and synthesis of next-generation high-performance structural materials, in particular, containing brittle covalent compounds.

## II. MATERIAL AND METHODS

*Sample fabrication*

TiC, VC, NbC, TaC, Mo$_2$C, WC powers, together with TiN$_{0.3}$, TiC$_{0.4}$, were utilized as starting materials for the synthesis of multi-principal compounds. Non-stoichiometric TiN$_{0.3}$ and TiC$_{0.4}$ were prepared through mechanical alloying method which has been details previously.[27,28] VC, NbC, TiC, TaC, WC, Mo$_2$C were purchased from Aladdin with a particle size between 1-3 μm, and purity of 99.5%. The raw materials were firstly ball-milled in a high energy planetary milling system (QM-3SP4, Nanjing, China) at 350 r min$^{-1}$ for 10 h under highly purity argon atmosphere. The milling instrument consisted of a main disk which could rotate at a speed of 100-600 rpm and accommodated two 500 mL WC vessels rotating in opposite directions so that the centrifugal forces alternately acted in the same and opposite directions. Each vessel contains numerous WC balls (diameters: 2-8 mm), and the ball to powder weight ratio (BPR) is selected as 10:1. Afterward, the mixtures were selected by an 80-mesh-sieve, and then pressed in a graphite die (inner diameter: 20 mm) at pressure of 40 MPa in a glove box under argon atmosphere. Subsequently, the compacts were sintered via SPS (SPS-3.20MK-IV, Sumitomo Coal Mining Company, Japan) under a vacuum of 10$^{-2}$ Pa between 1200-1700 ℃ for 10 min. The schematic diagrams of the spark plasma sintering (SPS) apparatus and die/punch/powder assembly are shown in Fig. S1 of the supplementary material. To minimize the radiation heat losses, the graphite die was surrounded with a 5.5 mm thick porous carbon felt insulation. The sintering temperature was measured using an infrared thermometer (IR-AH, GHINO, Japan) through a thermometer hole in the graphite die.

Heating rate of 100 °C min$^{-1}$ were selected, and a pressure of 40 MPa was applied on the compact throughout the sintering process. To assess the influence of non-stoichiometric TiN$_{0.3}$ or TiC$_{0.4}$ on the formation of multi-principal compounds, quasi-binary diffusion couple TiN$_{0.3}$-TMC (TMC: the equimolar mixing of TaC and NbC) was selected as a typical example. By contrast, the diffusion behavior of quasi-binary diffusion couple TiN-TMC was also discussed. Equimolar mixtures of TaC and NbC were mixed in a high energy planetary milling system at 300 r min$^{-1}$ for 2 h under highly purity argon atmosphere. Two groups of specimens were first equiaxially pressed into 10 mm diameter disks and then cold isostatically pressed under a pressure of 300 MPa. The quasi-binary diffusion couple were sintered via SPS at 1500 °C for 10 min. Samples for studying diffusion behavior were cut and the cross sections of the joints were polished for elemental analysis by an energy-dispersive X-ray spectroscope (EDX, APOLLO X, EDAX, America) from middle of the interface region. High-pressure and high-temperature (HPHT) experiments were also carried out at 5.5 GPa and 1400-1600 °C by using a hexahedron anvils press (CS-1B, Guilin, China). Sintering temperature was estimated by a W-Re5/26 thermocouple in the temperature of 700 °C to 1800 °C.

*Characterization methods*

After sintering, the specimens were characterized by X-ray diffraction (XRD) using a Rigaku diffractometer (D/MAX-2500/PC, Japan) with Cu Kα radiation (λ=0.15406 nm). The operating voltage and current were 40 KV and 200 mA. XRD results were recorded from 20 ° to 80 ° with a step size of 0.02 °. High energy synchrotron *x*-ray total scattering experiments were performed at beamline 11-ID-C of Advanced Photon Source (APS), Argonne National Laboratory. All the instrumental parameters were determined from the refinement of CeO$_2$ data (NIST standard) measured with the same X-ray energy. The X-ray wavelength is fixed at 0.11165 Å. The pair distribution function (PDF) was obtained by Fourier transformation from the high energy *x*-ray scattering data by using program PDFGetX2.

The analysis of microstructures and compositions were performed by a scanning electron microscope (SEM, FEI Scios, America) in backscattered electron (BSE) mode with an energy-dispersive X-ray spectroscope (EDX, Octane Plus-A, EDAX, America). Cross-sectional specimen for TEM and STEM observations was prepared by dual-beam focused ion beam (FIB) scanning microscopy (Versa 3D, FEI, USA) using Ga-ion beam and accelerating voltage ranging from 2 to 40 kV, followed by ion-milling (Gatan 691, Gatan, USA) operated from 1.5 to 0.5 kV and cooled with liquid nitrogen and then argon plasma cleaning (Solarus 950, Gatan, USA) to remove completely any residual amorphous film. TEM and STEM studies were carried out using a JEM-ARM300F (JEOL, Japan) microscope operating at 300 KV. The microscope is equipped with a spherical aberration corrector to enable sub-angstrom observation and dual EDS system for atomic resolution chemical mapping.

*Mechanical characterization*

Vickers hardness tests were performed using a diamond indenter (FM700, Future-Tech, Japan) at various loads in the range 50-1000g and at a dwell time of 10 s. Each hardness value represented an average of at least eight points taken on a mirror-like surface. Nano-indentation tests were conducted on *in-situ* nano-testing instrument (Hysitron Inc. Triboindenter, America) with the Berkovich indenter at room temperature. Each indentation experiment consisted of three steps with equal step time of 10 s: loading, holding the indenter at peak load, and unloading completely. The selected peak load is 8000 μN. Fused silica was used as reference sample for the initial tip calibration procedure. Data were processed using the Hysitron software providing load-displacement curves corrected for thermal drift and machine constants (frame compliance, transducer spring force, and electrostatic force constants). Load-displacement curves were analyzed according to the Oliver-Pharr method[29] by fitting a power law relationship to the unloading curve and determining the initial unloading stiffness. In order to keep the reliability of the experiment results, more than 100 indents were performed on each specimen.

Fracture toughness was determined through the Vickers indentation method proposed by Niihara[30] with a load of 5000 g and a holding time of 10 s.

For the Palmqvist cracks ($0.25 \leq l/a \leq 2.5$), the data can be described by

$$\left(\frac{K_{IC} \cdot \varphi}{Hv \cdot a^{0.5}}\right)\left(\frac{Hv}{E \cdot \varphi}\right)^{0.4} = 0.035\left(\frac{l}{a}\right)^{-0.5} \qquad (1)$$

and for the median cracks ($c/a \geq 2.5$), the corresponding expression is

$$\left(\frac{K_{IC} \cdot \varphi}{Hv \cdot a^{0.5}}\right)\left(\frac{Hv}{E \cdot \varphi}\right)^{0.4} = 0.129\left(\frac{c}{a}\right)^{-0.5} \qquad (2)$$

where $\varphi$ is the constraint factor ($\approx 3$), $K_{IC}$ is the fracture toughness, $E$ is the Young's modulus (taken from Nano-indentation measurement), $Hv$ is the Vickers hardness, $a$ is the length of half indentation diagonal, and $c$ is the crack length measured from the center of the indent, $l$ stands for Palmqvist crack length.

*Thermal characterization*

Thermal analysis was performed by thermal methods on a NETZSCH STA 449 F5 thermal analyzer under air atmosphere with the gas flow rate of 40 mL min$^{-1}$. The range of heating was from room temperature to 1550 ℃ with the heating rate of 10 ℃ min$^{-1}$. The samples were cut and polished into a disc shape of 6 mm in diameter and <1.5 mm in thickness for thermal diffusivity measurements. A thin layer of graphite was coated on the surface of sample disc to minimize errors from the emissivity of the material. The thermal diffusivity was then measured from room 27 ℃ to 577 ℃ using a laser flash method in TC9000 (Ulvac-Riko). The specimen densities were determined by Archimedes' method with distilled water as an immersing medium. And the temperature-dependent electrical resistivity was measured by a conventional four-probe method using a physical property measurement system (Quantum Design PPMS) for multi-principal compounds.

**III. Results and Discussion**

The multi-principal covalent compounds were prepared by solid-state spark plasma sintering (SPS) techniques. Transition metal carbides and nitrides were ball-milled in an argon environment and subsequently sintered at high temperature. Initially, TiN/VC/NbC with equimolar ratio was sintered at 1300 ℃, resulting in a mechanical mixture. Intriguingly, the single phase showed up when non-stoichiometric $TiN_{0.3}$ was used instead of TiN at 1300 ℃ (see Fig. S2 of the supplementary material). This temperature is much lower than previous two-component sintering, in which the single phase solid solutions were sintered at 2000 ℃ or above 2050 ℃, e.g. TaC-ZrC, TaC-HfC[22-24]. Increasing the number of components to four ($TiN_{0.3}$/VC/NbC/TiC), five ($TiN_{0.3}$/VC/NbC/TiC/TaC), six ($TiC_{0.4}$/VC/NbC/TiN/$Mo_2C$/WC) and seven ($TiN_{0.3}$/VC/NbC/TiC/TaC/$Mo_2C$/WC) covalent compounds, even with the addition of hexagonal WC and $Mo_2C$, all the products were surprisingly single phase, and also the single phases were favored at elevated temperature up to 1700 ℃ for more than four components, as shown in Fig. 1(a) and Fig. S3 of the supplementary material. These observations are consistent with previously studied high entropy alloys,[10-20] because of the high entropy of mixing at high temperature.

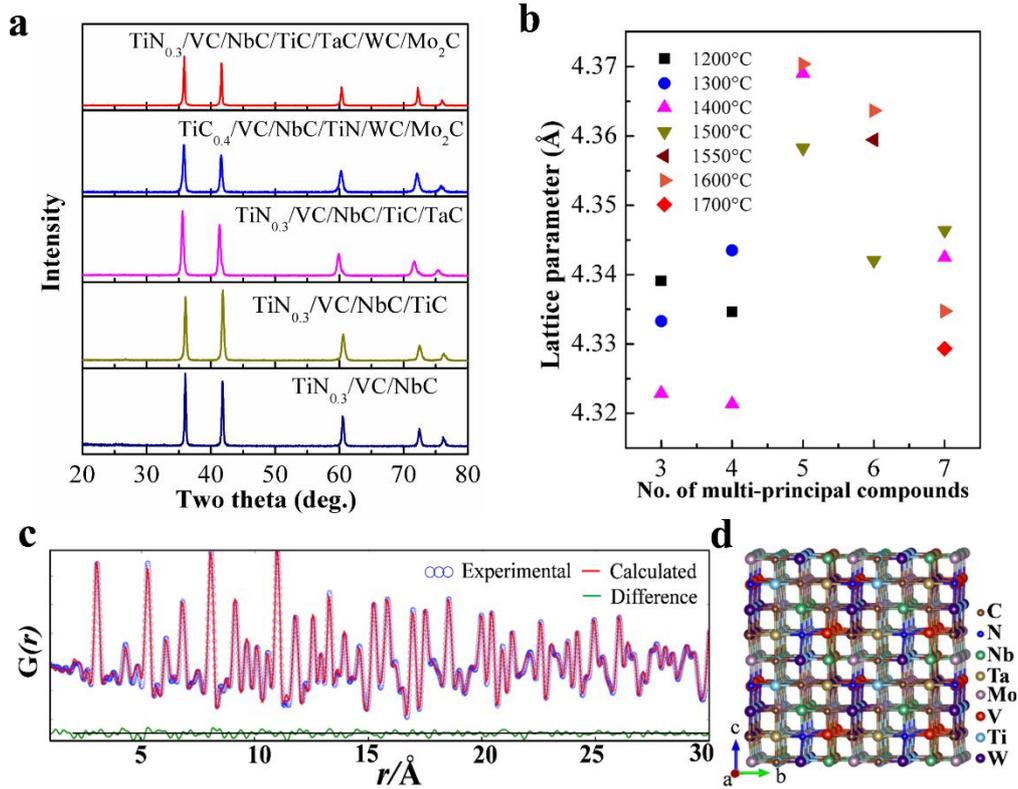

FIG. 1. XRD patterns and PDF analysis. (a) XRD data using Cu Kα radiation of all five MPTMCs after SPS (The details are presented in Fig. S3 of the supplementary material). (b) Lattice parameters of multi-principal compounds sintered at different temperatures. (c) PDF analysis for TiN$_{0.3}$/VC/NbC/TiC/TaC/Mo$_2$C/WC. (d) The FCC structural model of TiN$_{0.3}$/VC/NbC/TiC/TaC/Mo$_2$C/WC used for the PDF calculation.

Single phase formation of TiC$_{0.4}$/VC/NbC/TiN/Mo$_2$C/WC was also facilitated via high-pressure conditions at 5.5 GPa and above 1500 ℃. The compositions of hexagonal WC and Mo$_2$C from half- to equimolar ratios were also varied to confirm the single-phase feature for more than five components. The obtained powder X-ray diffraction (XRD) patterns of multi-principal covalent compounds, TiN$_{0.3}$/VC/NbC, TiN$_{0.3}$/VC/NbC/TiC, TiN$_{0.3}$/VC/NbC/TiC/TaC, TiC$_{0.4}$/VC/NbC/TiN/Mo$_2$C/WC, TiN$_{0.3}$/VC/NbC/TiC/TaC/Mo$_2$C/WC sintered at different temperatures can be indexed with a single FCC phase. The diffraction peaks shifted slightly with varying components at different temperatures. Broad reflections were observed for all the samples, revealing the high degree of disorder, similar to previous studied HEAs. However, the observed intensities of the Bragg diffraction peaks did not show a significant reduction from three to seven components. The lattice parameter for our seven-component compound was 4.338(1) Å for the sample sintered at 1500 °C, as shown in Fig. 1(b), similar to pure TiC (with a lattice parameter of 4.327 Å).

In order to further understand the atomic arrangement and local structural disorder in these compounds, the total scattering profiles are collected with high-energy synchrotron $x$-ray diffraction at APS. As a typical example, the obtained $x$-ray pair distribution function (PDF) spectra for $TiN_{0.3}$/VC/NbC/TiC/TaC/$Mo_2$C/WC is shown in Fig. 1(c). The observed PDF profile is well-fit by the one calculated using PDFgui[31] with an ideal FCC structural model and a random atomic arrangement. In the calculation, we use the lattice parameter obtained from the Rietveld refinement of high-energy synchrotron $x$-ray diffraction of the ideal structure, where the equilibrium position of each atom falls exactly on its assigned lattice site, but random [Fig. 1(d)]. The good match between experiment and calculation suggests that these chemically disordered compounds are in a well-defined single phase with negligible lattice distortion.

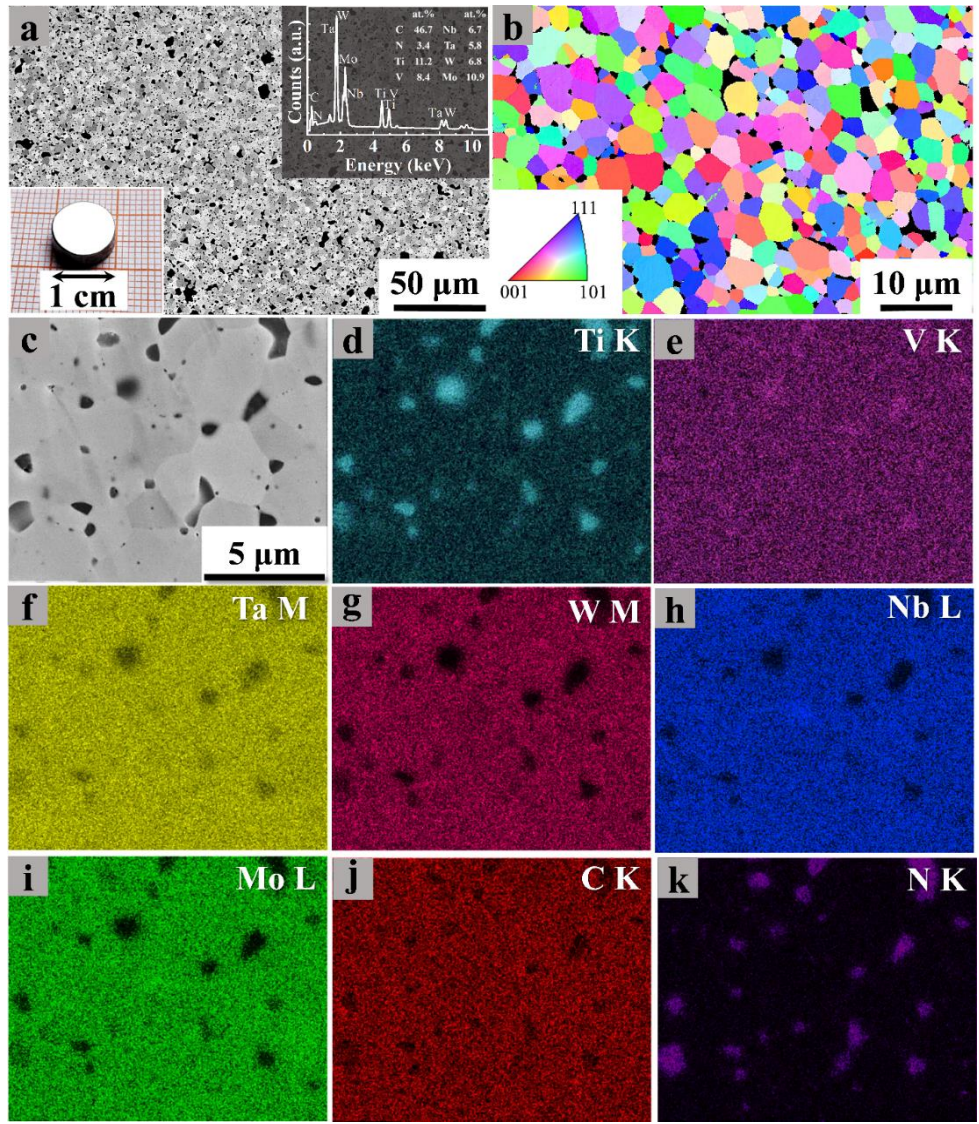

FIG. 2. Microstructure of the TiN$_{0.3}$/VC/NbC/TiC/TaC/Mo$_2$C/WC multi-principal transition metal carbides synthesized at 1600 ℃. (a) The BSE image reveals uniform grain size; the composition is approximately consistent with the desired results, as shown from the EDX spectroscopy inset. (b) IPF map. (c) Magnified BSE image obtained from the region in (a). (d-k) Corresponding SEM-EDS maps of (c) revealing a random distribution of (d) Ti, (e) V, (f) Ta, (g) W, (h) Nb, (i) Mo, (j) C and (k) N elements.

Microstructure characterization of the single phase synthesized from TiN$_{0.3}$/VC/NbC/TiC/TaC/Mo$_2$C/WC at 1600 ℃ is illustrated in Fig. 2 which shows a scanning backscattered electron (BSE) image and electron backscatter diffraction (EBSD) map collected in a scanning electron microscope (SEM). EBSD reveals a fully crystallized and faceted morphology with an average grain size of approximately 1.8 μm, consistent with estimation from Scherrer formula (about 2 μm). The analysis of crystallographic misorientation reveals that the high angle boundaries (with a fraction of 90.7%) are predominant across the grain boundaries. The misorientation between neighboring grains may be a factor for boosting carbide/nitride diffusion and could therefore facilitate the growth of multi-principal compounds.[32,33] We also find that increasing the sintering temperature tends to promote an increase in the grain size, with the analysis of EBSD microstructure and Scherrer estimates giving an average grain size of ~0.5 μm at 1300 °C and >10 μm at 1700 °C (Fig. S4 of the supplementary material). Compositional analysis via energy-dispersive *x*-ray spectroscopy (EDS) finds that the sample composition is close to the starting mixtures, single-phase (TiVNbTaMoW)C is formed for the majority of crystal grains, and $TiN_{0.3}$ is concentrated along with a very small contribution of the equiatomic metal carbides at grain boundaries.

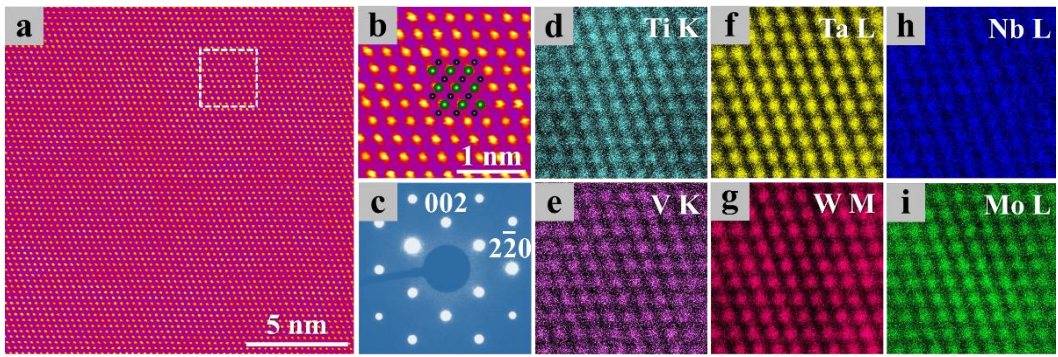

FIG. 3. Atomic structure of $TiN_{0.3}$/VC/NbC/TiC/TaC/Mo$_2$C/WC sintered at 1600 °C. (a) High-angle annular dark-filed (HAADF) image showing the atom structure projected along (110) zone-axis. (b) Magnified HAADF image obtained from the marked square region in (a), with inset labeled by the atom projection of TiC along (110) zone-axis (Ti: Teal, C: Black). (c) Selected area electron diffraction (SAED) pattern taken along (110) zone-axis. (d-i) Corresponding STEM-EDS maps of (b) revealing a random distribution of (d) Ti, (e) V, (f) Ta, (g) W, (h) Nb and (i) Mo atoms.

To gain more insight into the atomic structure of MPTMC, high-resolution scanning transmission electron microscopy (STEM) analysis was performed. High-angle annular dark-field (HAADF) imaging shows clearly an ordered structure with uniform contrast of all atom columns, as shown in Fig. 3(a). Note that only the heavier cation atoms of Ti/V/Ta/W/Nb/Mo show bright contrast in the Z-contrast (where Z refers to atomic number) HAADF image, while the C atoms are barely visible. As shown in the magnified HAADF image in Fig. 3(b), the atomic structure fits well the face-centered cubic lattice symmetry of TiC (Fm-3m), with all cations sitting on the Ti site. Furthermore, selected area electron diffraction (SAED) reveals the absence of any superlattice reflection [Fig. 3(c)] from reciprocal space. The result is indicative of a superlattice-ordering-free atomic structure for the MPTMCs. In addition to structural observations, scanning transmission electron microscopy with energy dispersive X-ray spectroscopy (STEM-EDS) is used to analyze structure and chemistry on the local scale. Fig. 3(d-i) shows the intensity maps for the characteristic EDS signals of Ti, V, Ta, W, Nb and Mo, respectively. It is clearly observed that each atom column contains all cation elements, while appears free from any atom segregation. The atomic level STEM results provide direct evidence of a random distribution of all cation atoms in the MPTMCs. The random distribution of all cations suggests that a mutual diffusion process occurred in these carbides. This can also be understood by the quasi-binary diffusion couples of $TiN_{0.3}$-TMCs and TiN-TMCs (TMCs, the mixing of equimolar TaC and NbC) from the metallographic observations of interface area. Elemental analysis of two couple samples shows that the diffusion of Ta and Nb in $TiN_{0.3}$-TMCs is about four times faster than those in TiN-TMCs (Fig. S5 and Table S2 of the supplementary material). This rapid interfacial diffusion in $TiN_{0.3}$-TMCs with the introduction of non-stoichiometric $TiN_{0.3}$ greatly promotes the formation of single-phase solid solutions when the sintered temperature was elevated.

To assess the mechanical properties of these compounds, Vickers hardness was measured on the polished surface for the microhardness tester. Fracture toughness was determined through the Vickers indentation method proposed by Niihara[30]. Obtained Vickers hardness and fracture toughness results for the multi-principal compounds are shown in Fig. 4(a) and Fig. S6 of the supplementary material, $TiN_{0.3}$/VC/NbC/TiC sintered at different temperatures showed the highest Vickers hardness values in the range of 23.0-26.0 GPa, close to that of the $TiN_{0.3}$/VC/NbC/TiC/TaC samples with a Vickers hardness of 23.0 GPa. Significantly, $TiN_{0.3}$/VC/NbC/TiC/TaC/$Mo_2C$/WC sintered at 1300 °C showed the highest fracture toughness of 8.4 MPa·m$^{1/2}$. Generally, the trade-off between hardness and toughness of obtained multi-principal compounds are significantly higher than the highest values for individual refractory metal carbides and composites of 7.8 MPa·m$^{1/2}$, alumina matrix composites[34-43] of 5.8 MPa·m$^{1/2}$ and mullite and its composites[44-49] of 4.0 MPa·m$^{1/2}$. We also found that a high sintering temperature tends to decrease the Vickers hardness and fracture toughness over 1600 °C, which is likely caused by rapid grain growth. SEM images of fracture surfaces for $TiN_{0.3}$/VC/NbC/TiC/TaC/$Mo_2C$/WC shown in Fig. S4 of the supplementary material reveal that at the early stage, a uniform microstructure is formed at temperatures of 1300 and 1400 °C; the trans-granular fracture become significant when sintered at 1700 °C. Furthermore, the thermal stability of two representative samples of $TiN_{0.3}$/VC/NbC/TaC/$Mo_2C$ and $TiN_{0.3}$/VC/NbC/TiC/TaC/$Mo_2C$/WC is evaluated by thermogravimetry curves measured in air. At a heating rate of 10 °C min$^{-1}$, the onset oxidation temperature of $TiN_{0.3}$/VC/NbC/TiC/TaC/$Mo_2C$/WC is found to be 882 °C (Fig. S7 of the supplementary material), higher than those of WC-Co (600-800 °C)[50] and nano-crystalline diamond (677 °C)[51].

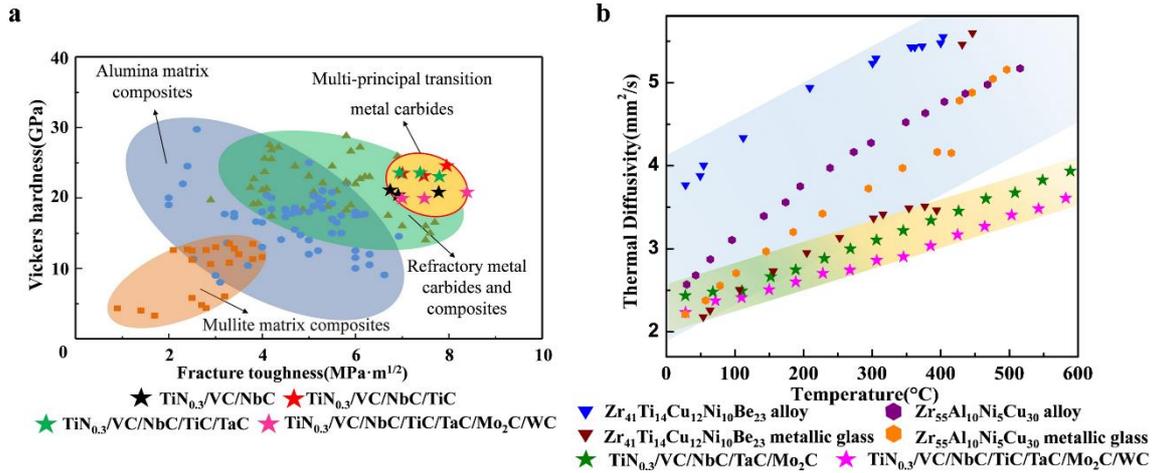

FIG. 4. Comparison of mechanical and thermal properties. (a) The distributions of fracture toughness-Vickers hardness of the multi-principal carbides synthesized here, comparable to those of alumina matrix composites[34-43], mullite matrix composites[44-49] and refractory metal carbides and composites[52-64]. (b) Temperature dependence of thermal diffusivity for multi-principal carbides and metallic glasses [65,66].

Another important feature of the present multi-principal compounds is the low glass-like thermal diffusivity observed over the broad temperature range [Fig. 4(b) and Fig. S8 of the supplementary material]. The thermal diffusivity for five and seven multi-principal compounds was measured using the laser-flash method as a function of temperature (in the range 27 ℃-600 ℃). Interestingly, our thermal-diffusivity results (2.23 and 2.43 mm$^2$ s$^{-1}$) for five and seven components are close to those of ZrTiCuNiBe (2.18 mm$^2$ s$^{-1}$)[65] and ZrAlNbCu (2.17 mm$^2$ s$^{-1}$) metallic glass[66]. The observed thermal diffusivity of these two samples does not exceeding that of ZrTiCuNiBe (2.18 mm$^2$ s$^{-1}$) metallic glass in the temperature range of 27-377 ℃, and is much smaller than those of typical ceramics and their composites, e.g. ZrC and ZrB$_2$-based high temperature ceramic composites[67,68]. Such low thermal diffusivities are typically found among electrically insulating amorphous solids and glasses. The multi-principal compounds display much lower thermal diffusivity compared to TiC (ideal TiC, 7.53 mm$^2$ s$^{-1}$ at 25 ℃)[69] with the same crystal symmetry, which can be attributed to its striking difference with complex random compositions and statistical disorder in these multi-principal compounds. Furthermore, temperature dependent resistivity of these five-and seven-principal compounds showed metallic behavior without superconductivity, different from the constituent compounds.

**IV. Conclusion**

In summary, our findings provide a new multi-principal design strategy for covalent transition metal carbides by controllable interfacial diffusion, demonstrating that the combination of specific hardness and toughness accessible is better than previously thought, and increasing the damage tolerance for structural applications. Furthermore, the attractive combination of physical and mechanical properties in the multi-principal covalent compounds described here is obtainable by simple mechanical alloying. These results represent a starting point for exploration of a large family of multi-principal transition-metal carbides/nitrides. The desirable mechanical, thermal, and physical properties of these new multi-principal compounds make them attractive candidates for a broad range of technological applications.

**SUPPLEMENTARY MATERIAL**

See the supplementary material for details of experimental details and additional figures.

**Acknowledgements**

Author contributions: C.P., X.G. and M.Z.W. contributed equally. M.Z.W. and H.Y.G. conceived and designed the project, C.P., Q.Y., D.X.T., H.T., Q.Z., Y.C.Z. performed the experiments and characterizations. X.G., X.M.L and X.D.B conducted the TEM and analyzed the results. L.L.W. performed nano-indentation tests and DSC analysis. Y.R. and F.X.Z collected and analyzed PDF data, Y.H.W. and B.Z. performed SEM observations. Q.ZH. W.G.Y., H.K.M performed the thermal diffusivity measurements and analyzed the data. B.G. and H.X. conducted resistivity measurements. All authors analyzed the data and discussed the results. H.Y.G., M.Z.W., X.G. and W.M prepared the manuscript with the contributions of all authors. This work was supported by the Natural Science Foundation of Hebei Province (E2016203425 and E2015203232); China Postdoctoral Special Funding (2015T80895); Science and Technology Research Youth Fund of Colleges and University of Hebei Province (QN20131092); Heavy Machinery Synergy Innovation (ZX01-20140100-01) and Hebei Province First Batch of Young Talent Support Plan (the second support cycle, No.(2016)9). H.G. acknowledges the financial supports from the National Natural Science Foundation of China (NSFC) under Grants No. 51201148, No. U1530402 and the Thousand Youth Talents Plan. W.L.M. is supported by the Department of Energy through the Stanford Institute for Materials and Energy Sciences DE-AC02-76SF00515.

# Supplementary material

# Diffusion-controlled Alloying of Single-phase Multi-principal Covalent Transition Metal Carbides with Enhanced Damage tolerance and Exceptional Thermal Properties

Chong Peng[1,2,a], Xiang Gao[2,a], Mingzhi Wang[1,a,b], Lailei Wu[1], Hu Tang[1,2], Xiaoming Li[3], Qian Zhang[2], Yang Ren[4], Fuxiang Zhang[5], Yuhui Wang[6], Bing Zhang[1], Bo Gao[2], Qin Zou[1], Yucheng Zhao[1], Qian Yang[1], Dongxia Tian[1], Hong Xiao[2], Huiyang Gou[2,b], Wenge Yang[2], Xuedong Bai[3], Wendy L. Mao[7] and Ho-kwang Mao[2,8]

[1]*State Key Laboratory of Metastable Materials Science and Technology, Yanshan University, Qinhuangdao 066004, People's Republic of China;*

[2]*Center for High Pressure Science and Technology Advanced Research, Beijing 100190, People's Republic of China;*

[3]*Beijing National Laboratory for Condensed Matter of Physics and Institute of Physics, Chinese Academy of Sciences, Beijing 100190, People's Republic of China;*

[4]*Advanced Photon Source, Argonne National Laboratory, Argonne, Illinois 60439, USA;*

[5]*Division of Materials Science & Technology, Oak Ridge National Laboratory, Oak Ridge TN 37831, USA;*

[6]*National Engineering Research Center for Equipment and Technology of Cold Strip Rolling, Yanshan University, Qinhuangdao 066004, China;*

[7]*Geological Sciences, Stanford University, Stanford, CA 94305;*

[8]*Geophysical Laboratory, Carnegie Institution of Washington, Washington, DC 20015, USA*

[a]C.P., X.G. and M.Z.W. contributed equally to this work.

[b]Authors to whom correspondence should be addressed. Electronic addresses: wmzw@ysu.edu.cn(M.Z.W.); huiyang.gou@hpstar.ac.cn(H.Y.G.)

Table S1. Nanoindentation hardness (GPa) and Young's modulus (GPa) of MPTMCs--TiN$_{0.3}$/VC/NbC/TaC/Mo$_2$C and TiN$_{0.3}$/VC/NbC/TiC/TaC/Mo$_2$C/WC (Statistical data are shown as mean ± standard deviation.)

| Samples | Nanoindentation hardness/GPa | Elastic modulus/GPa |
|---|---|---|
| 5 | 28.0 ± 2.4 | 519.2 ± 28.8 |
| 7 | 28.0 ± 1.6 | 554.9 ± 28.3 |

Table S2. The measured thickness of diffusion layer of Ti, Nb, Ta by the technique of quasi-binary diffusion couples of TiN$_{0.3}$-TMC and TiN-TMC sintered at 1500 °C.

| Quasi-binary diffusion couples | Thickness of diffusion layer (µm) | | |
|---|---|---|---|
| | Ti | Nb | Ta |
| **TiN-TMC** | 11~12 | 2~3 | 4~5 |
| **TiN$_{0.3}$-TMC** | 21~22 | 12~13 | 16~17 |

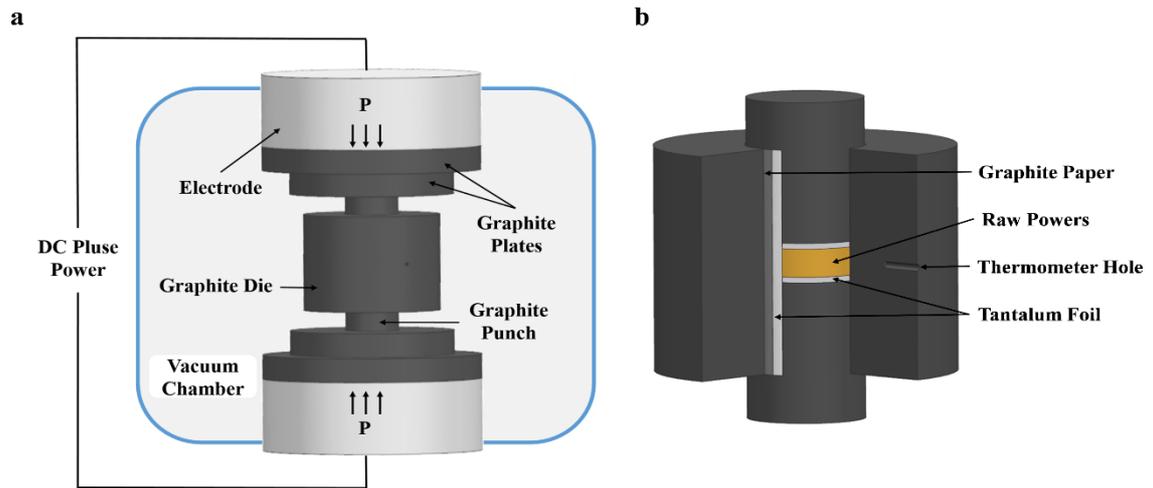

Figure S1. Schematic diagrams of (a) the spark plasma sintering (SPS) apparatus and (b) die/punch/powder assembly for SPS.

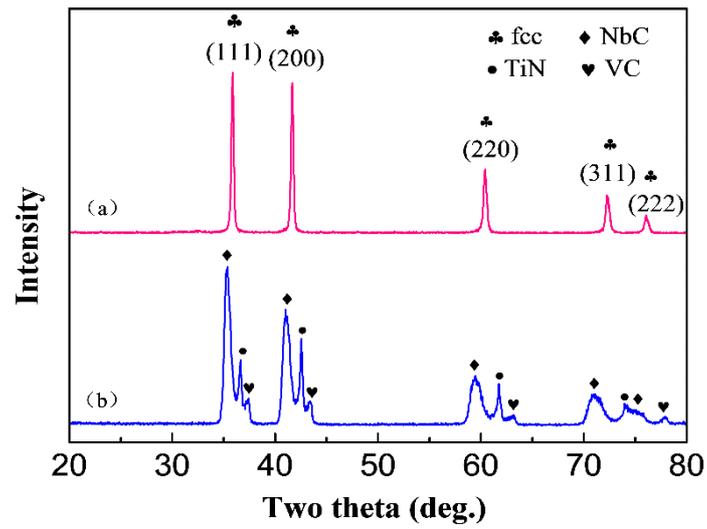

Figure S2. XRD patterns of (a) TiN$_{0.3}$/VC/NbC and (b) TiN/VC/NbC sintered at 1300 ℃.

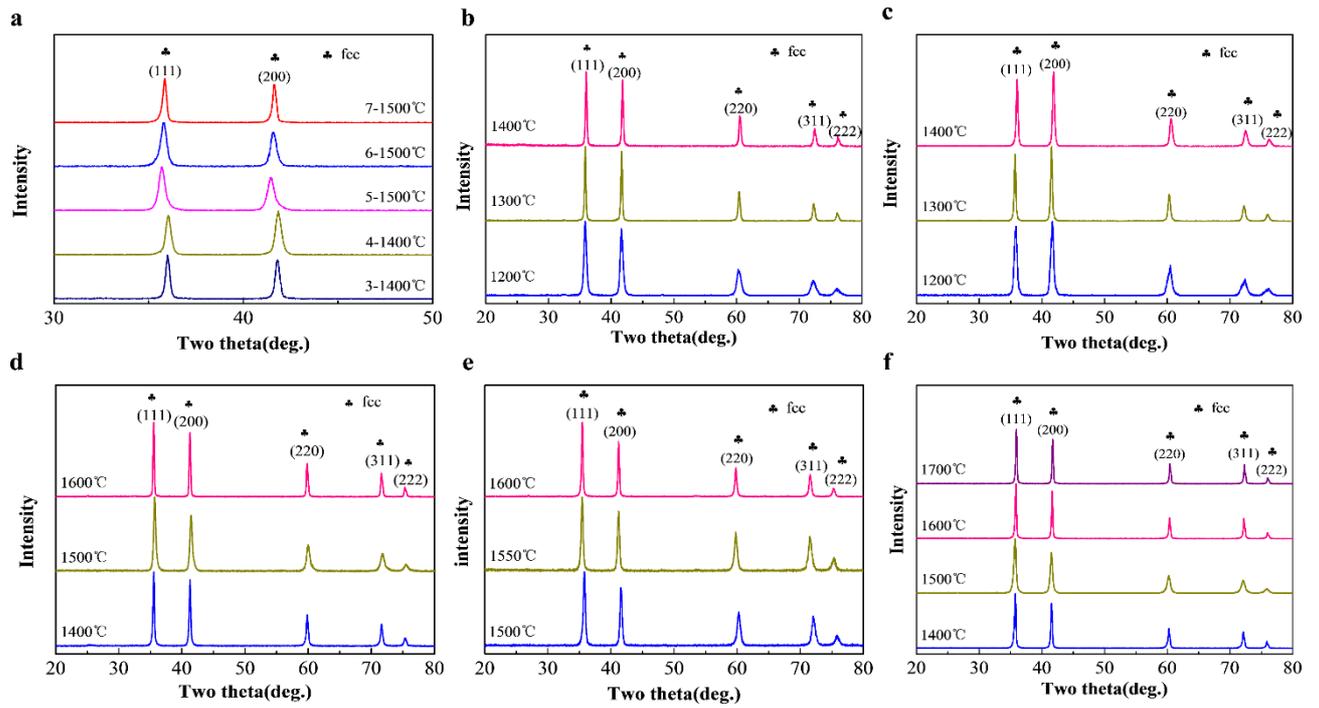

Figure S3. XRD patterns of MPTMCs sintered at different temperatures. (a) XRD data of MPTMCs with different components at desired temperatures. (b) TiN$_{0.3}$/VC/NbC. (c) TiN$_{0.3}$/VC/NbC/TiC. (d) TiN$_{0.3}$/VC/NbC/TiC/TaC. (e) TiC$_{0.4}$/VC/NbC/TiN/Mo$_2$C/WC. (f) TiN$_{0.3}$/VC/NbC/TiC/TaC/Mo$_2$C/WC.

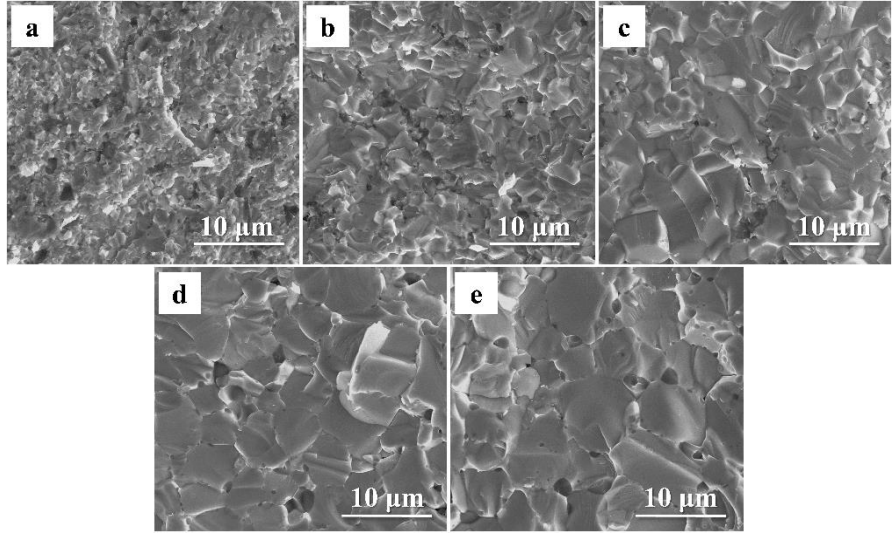

Figure S4. SEM images of fracture surfaces of TiN$_{0.3}$/VC/NbC/TiC/TaC/Mo$_2$C/WC sintered at different temperatures. (a) 1300 ℃. (b) 1400 ℃. (c) 1500 ℃. (d) 1600 ℃. (e) 1700 ℃.

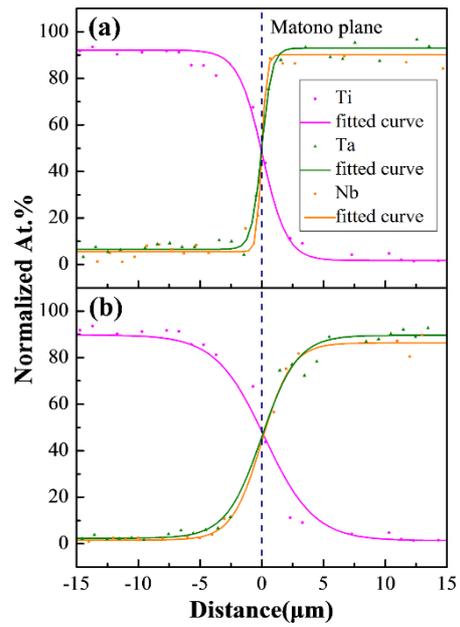

Figure S5. Atomic distributions of Ti, Ta, and Nb atoms of (a) TiN-TMC, (b) TiN$_{0.3}$-TMC couple samples after sintered at 1500 °C for 10 min.

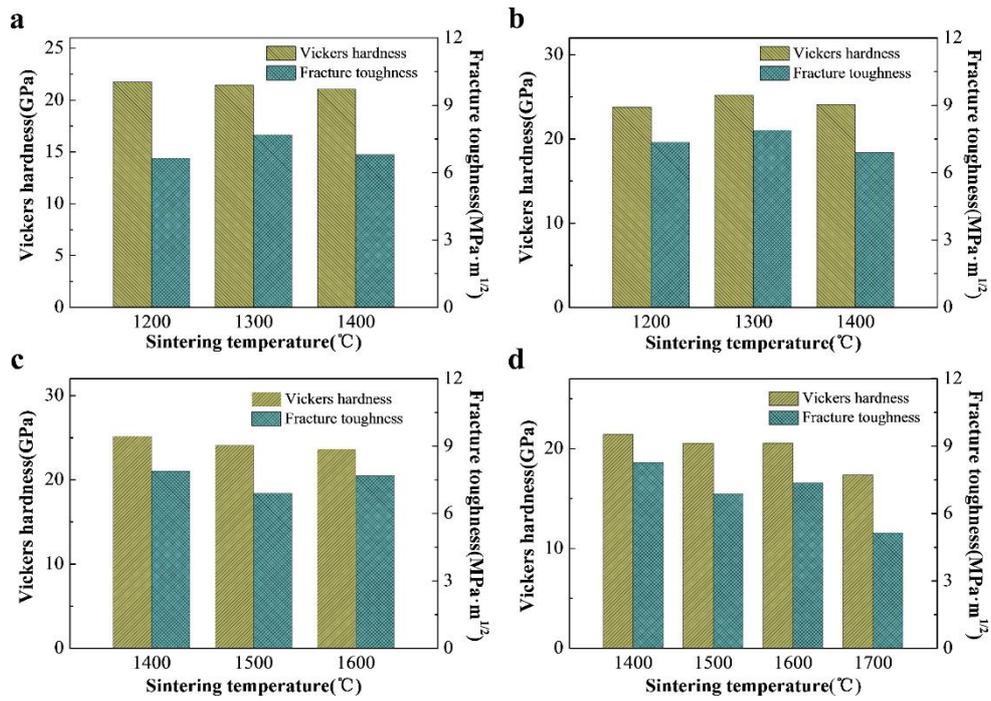

Figure S6. Vickers hardness and fracture toughness of MPTMCs with different sintering temperatures. (a) TiN$_{0.3}$/VC/NbC. (b) TiN$_{0.3}$/VC/NbC/TiC. (c) TiN$_{0.3}$/VC/NbC/TiC/TaC. (d) TiN$_{0.3}$/VC/NbC/TiC/TaC/Mo$_2$C/WC.

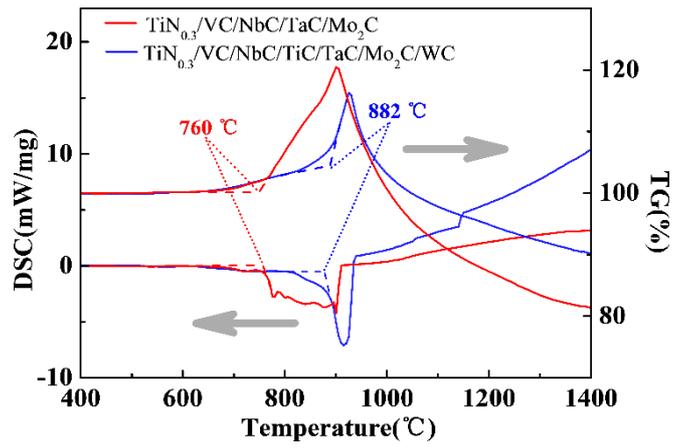

Figure S7. DSC-TG analysis of $TiN_{0.3}$/VC/NbC/TaC/$Mo_2$C and $TiN_{0.3}$/VC/NbC/TiC/TaC/$Mo_2$C/WC in air with the temperatures of 25-1550 ℃.

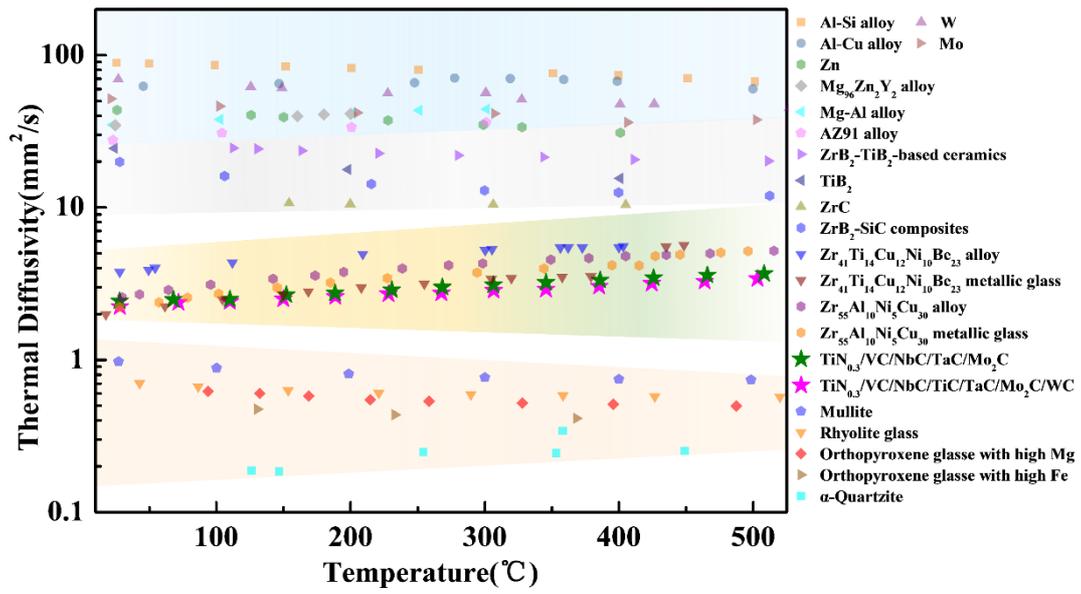

Figure S8. Thermal diffusivity as a function of temperature of MPTMCs (pink stars and green stars) in comparison with available data on known metals and ceramics.(4-15)